# Resonance in the electron-doped high-$T_c$ superconductor Pr$_{0.88}$LaCe$_{0.12}$CuO$_{4-\delta}$

Stephen D. Wilson [1], Pengcheng Dai[1,2], Shiliang Li[1], Songxue Chi[1], H. J. Kang[3,4], and J. W. Lynn[3]

[1] Department of Physics and Astronomy, The University of Tennessee, Knoxville, Tennessee 37996-1200, USA

[2] Center for Neutron Scattering, Oak Ridge National Laboratory, Oak Ridge, Tennessee 37831, USA

[3] NIST Center for Neutron Research, National Institute of Standards and Technology, Gaithersburg, Maryland 20899-8562 USA

[4] Departmenth of Materials Science and Engineering, University of Maryland, College Park, Maryland 20742 USA

**In conventional superconductors, the interaction that pairs the electrons to form the superconducting state is mediated by lattice vibrations (phonons)[1]. In high-transition temperature (high-$T_c$) copper oxides, it is generally believed that magnetic excitations play a fundamental role in the superconducting mechanism because superconductivity occurs when mobile 'electrons' or 'holes' are doped into the antiferromagnetic parent compounds[2]. Indeed, a sharp magnetic excitation termed "resonance" has been observed by neutron scattering in a number of hole-doped materials[3-11]. The resonance is intimately related to superconductivity[12], and its interaction with charged quasi-particles observed by photoemission[13,14], optical conductivity[15], and tunneling[16] suggests that it plays a similar role as phonons in**

conventional superconductors. However, the relevance of the resonance to high-$T_c$ superconductivity has been in doubt because so far it has been found only in hole-doped materials[17]. Here we report the discovery of the resonance in electron-doped superconducting $Pr_{0.88}LaCe_{0.12}CuO_{4-\delta}$ ($T_c$ = 24 K). We find that the resonance energy ($E_r$) is proportional to $T_c$ via $E_r = 5.8 k_B T_c$ ($k_B$ is the Boltzmann's constant) for all high-$T_c$ superconductors irrespective of electron- or hole-doping (Fig. 1e). Our results demonstrate that the resonance is a fundamental property of the superconducting copper oxides and therefore must play an essential role in the mechanism of superconductivity.

Although the interaction of electrons with phonons or magnetic excitations can cause electron pairing and superconductivity, we focus on magnetic excitations because the resonance is intimated related to superconductivity and also present in several classes of hole-doped high-$T_c$ materials. The resonance is a sharp magnetic excitation centered at the wavevector $\mathbf{Q} = (1/2, 1/2)$ in the two-dimensional reciprocal space of the $CuO_2$ planes, which corresponds to the antiferromagnetic (AF) Bragg position of the undoped compounds (Fig. 1a). It was first discovered in the hole-doped bilayer (each lattice unit cell has two $CuO_2$ planes) high-$T_c$ superconductor $YBa_2Cu_3O_{6+x}$ (YBCO)[3]. Its intensity grows below $T_c$ and its energy ($\hbar\omega$) scales approximately with $k_B T_c$ (Fig. 1e)[4-8]. At energies below the resonance, spin fluctuations peak at incommensurate wavevectors[8] and disperse inwards toward the resonance[18,19]. Such behavior is remarkably similar to that of the hole-doped $La_{2-x}(Ba,Sr)_xCuO_4$[20-22]. Although the resonance has also been observed in the hole-doped bilayer $Bi_2Sr_2CaCu_2O_{8+\delta}$ [Bi(2212)][9,10] and in the single layer

$Tl_2Ba_2CuO_{6+\delta}$[11], the wavevector and energy dependence of the mode has been determined only for YBCO[3-8] because of the small available single crystal volumes in the Bi(2212)[9,10] and $Tl_2Ba_2CuO_{6+\delta}$[11]. Therefore, it has not been possible to directly and systematically compare the neutron results with those of the photoemission, which are obtained mostly for Bi(2212)[13,14].

We are studying the magnetic excitations of electron-doped materials to understand the electron-hole symmetry in high-$T_c$ superconductors. We grew large single crystals of $Pr_{0.88}LaCe_{0.12}CuO_{4-\delta}$ (PLCCO) using the traveling solvent floating zone technique and annealed the samples to obtain optimal superconductivity with $T_c =24$ K (Fig. 1a)[23,24]. PLCCO is a single-layer electron-doped copper oxide and was chosen to avoid the static AF order that coexists with superconductivity in $Nd_{1.85}Ce_{0.15}CuO_4$ (NCCO)[25]; the $T_c = 24$ K PLCCO is phase pure without static AF order[23,26]. Our elastic **Q**-scans through the expected AF Bragg positions confirm no static AF order to at least 600 mK in our PLCCO samples (Figs. 1b and 1c).

We first probe the low-energy magnetic excitations of PLCCO using the SPINS cold neutron triple-axis spectrometer. The magnetic excitations are commensurate and centered around **Q** = (1/2,1/2,0) at all temperatures (Figs. 2a-c) similar to that of NCCO[25]. They can be well described by Gaussians on linear backgrounds and are not resolution-limited. Fourier transforms of the Gaussian peaks in Figs. 2a-c give the dynamic spin correlation lengths of $\xi \approx 96 \pm 15$, $80 \pm 10$, and $94 \pm 24$ Å for $\hbar\omega= 0.5$, 1.5, and 3.5 meV, respectively. With increasing temperature, the scattering above background is virtually unchanged from 2 K ($T_c$ - 22 K) to 30 K ($T_c$ + 6 K) but decreases slightly at 55 K due to increased background (Figs. 2a-c). The temperature dependence of the scattering at $\hbar\omega$ = 1.5 meV shows no observable anomaly across $T_c$ (Fig. 2d), thus suggesting that the magnetic excitations are gapless and different from that of NCCO[25]. The energy scans at **Q** = (1/2, 1/2, 0) confirm that the magnetic scattering between 0.5 meV and 4.5 meV is virtually temperature independent between 2 K and 30 K (Fig. 2e), and therefore does not follow the population factor $1/(1-e^{-\hbar\omega/k_B T})$ expected for simple bosonic excitations.

To study the magnetic excitations for energies above 4.5 meV, we use the HB-1 and BT-9 thermal neutron triple-axis spectrometers. Figures 3a-c show **Q**-scans through (1/2, 1/2, 0) for $\hbar\omega$ = 3.5, 8.0, and 10 meV at $T$ = 2, 30, and 80 K. In contrast to the temperature independent low-energy ($\hbar\omega \leq$ 4.5 meV) magnetic excitations below $T_c$ (Fig. 2), the integrated intensity above background around (1/2, 1/2, 0) at $\hbar\omega$ = 8 and 10 meV shows a significant enhancement on cooling from 30 K to 2 K, but hardly changes on warming from 30 K to 80 K (Fig. 4d). The **Q**-width of the peak at $\hbar\omega$ = 10 meV is temperature independent and resolution-limited, giving a minimum $\xi \approx$ 45 ± 5 Å (Fig. 3c). Similar scans using better collimations on BT-9 (Fig. 4e) are again resolution-limited, giving $\xi \approx$ 51 ± 7 Å. Figure 3d shows energy scans at the peak center [**Q** = (1/2, 1/2, 0)] compared to background [**Q** = (0.5875, 0.5875, 0)] positions for temperatures 2, 30, and 80 K. The weak, dispersion-less peak at $\hbar\omega$ = 6 meV and the gradual rising background scattering with increasing energy are due to the $Pr^{3+}$ crystalline electric-field (CEF) excitations in the tetragonal unit cell of PLCCO[27]. Its intensity simply renders a shift in the background scattering on which the sharply localized magnetic excitations at **Q** = (1/2, 1/2, 0) rest (see Supplementary Information). The temperature difference spectrum (2 K - 30 K) shows a clear intensity gain around ~11 meV at **Q** = (1/2, 1/2, 0) (Fig. 3e).

Figure 4a shows the energy dependence of the scattering at **Q** = (1/2, 1/2, 0) and background (0.6, 0.6, 0) positions using BT-9. The scattering at (1/2, 1/2, 0) around ~11 meV is systematically higher below $T_c$ while the background intensity at (0.6, 0.6, 0) is temperature-independent between 2 K and 30 K. The temperature difference spectrum below and above $T_c$ (2 K - 30 K) in Fig. 4c shows a clear resolution-limited resonance peak centered at $\hbar\omega \approx 11$ meV, consistent with Fig. 3e. A constant energy scan at $\hbar\omega = 10$ meV confirms that the peak is centered at (1/2, 1/2, 0) (Fig. 4b) while a similar scan at $\hbar\omega = 15$ meV shows only background scattering (Fig. 4b). Finally, in Fig. 4d we plot the temperature dependence of the scattering at (1/2, 1/2, 0) and the integrated intensity around (1/2, 1/2, 0) above the background for $\hbar\omega = 10$ meV. They both increase dramatically below the onset of $T_c$ and are remarkably similar to that of the optimally doped YBCO[3,4] and Bi(2212)[9].

To summarize the neutron scattering results in Figs. 2-4, we plot the dispersion of the observed magnetic excitations in Fig. 1d and the resonance energy as a function of $T_c$ in Fig. 1e[3-12]. While the magnetic excitations are broader than the instrumental resolution for $\hbar\omega \leq 3.5$ meV and resolution-limited for $\hbar\omega \geq 4.5$ meV, they are commensurate and centered at (1/2, 1/2, 0) at all measured energies. This differs from the hole-doped materials, where incommensurate spin fluctuations below the resonance merge into it[18-21].
On the other hand, the resonance energy of 11 meV ($E_r \approx 5.3\, k_B T_c$) for the PLCCO is remarkably close to the universal value of $E_r \cong 5.8\, k_B T_c$ for all materials (Fig. 1e)[3-12].

Our results reveal several important conclusions for the electron-hole symmetry of the magnetic excitations in high-$T_c$ copper oxides. First and foremost, the discovery of the magnetic resonance in electron-doped PLCCO with $E_r \approx 5.3\, k_B T_c$ suggests that the resonance is a common feature for high-$T_c$ superconductors irrespective of electron- or hole-doping. Second, the observation of commensurate spin fluctuations below the resonance (Fig. 1d) implies that the intimate connection between incommensurate spin fluctuations and the resonance in hole-doped materials is not a universal feature[18-21]. At present, it is unclear how stripe models can account for these differences[20,28]. Third, the magnetic excitations ($0.5 \le \hbar\omega \le 16$ meV) in the electron-doped PLCCO are gapless, decrease monotonically with increasing energy, and are virtually temperature independent between $2 \le T \le 30$ K except for the appearance of the resonance below $T_c$ (Figs. 2-4). While such behavior differs from optimally hole-doped YBCO[3-6] and La$_{2-x}$Sr$_x$CuO$_4$[21], the temperature independent low-energy ($0.5 \le \hbar\omega \le 4.5$ meV) magnetic scattering is remarkably similar to the quantum critical scattering in heavy Fermion UCu$_{5-x}$Pd$_x$[29]. Finally, the discovery of the magnetic resonance in electron-doped PLCCO, where charged quasiparticles can also be probed by photoemission[30], allows a systematic comparison of their properties in the same bulk sample hitherto not possible for any other high-$T_c$ superconductors. This should open new avenues of research to understand the exotic properties of high-$T_c$ copper oxides.

1. Bardeen, J., Cooper, L. N., & Schrieffer, J. R. Theory of Superconductivity. *Phys. Rev.* **108**, 1175-1204 (1957).

2. Orenstein, J. & Millis, A. J. Advances in the Physics of High-Temperature Superconductivity. *Science* **288**, 468-474 (2000).

3. Rossat-Mignod, J. *et al.* Neutron scattering study of the $YBa_2Cu_3O_{6+x}$ system. *Physica C* **185**, 86-92 (1991).

4. Mook, H. A. *et al.* Polarized neutron determination of the magnetic excitations in $YBa_2Cu_3O_7$. *Phys. Rev. Lett.* **70**, 3490-3493 (1993).

5. Fong, H. F. *et al.* Spin susceptibility in underdoped $YBa_2Cu_3O_{6+x}$. *Phys. Rev. B* **61**, 14773-14786 (2000).

6. Dai, P., Mook, H. A., Hunt, R. D., & Doğan, F. Evolution of the resonance and incommensurate spin fluctuations in superconducting $YBa_2Cu_3O_{6+x}$. *Phys. Rev. B* **63**, 054525 (2001).

7. Stock, C. *et al.* Dynamic stripes and resonance in the superconducting and normal phases of $YBa_2Cu_3O_{6.5}$ ortho-II superconductor. *Phys. Rev. B* **69**, 014502 (2004).

8. Hayden, S. M., Mook, H. A., Dai, P., Perring, T. G., Doğan, F. The structure of the high-energy spin excitations in a high-transition-temperature superconductor. *Nature* **429**, 531-534 (2004).

9. Fong, H. F. *et al.* Neutron scattering from magnetic excitations in $Bi_2Sr_2CaCu_2O_{8+\delta}$. *Nature* **398**, 588-591 (1999).

10. He, H. *et al.* Resonant spin excitations in a overdoped high temperature superconductor. *Phys. Rev. Lett.* **86**, 1610-1613 (2001).

11. He, H. *et al.* Magnetic resonant mode in the single-layer high-temperature superconductor $Tl_2Ba_2CuO_{6+\delta}$. *Science* **295**, 1045-1047 (2002).

12. Dai, P. *et al.* The magnetic excitation spectrum and thermodynamics of high-$T_c$ superconductors. *Science* **284**, 1344-1347 (1999).

13. Damascelli, A., Hussain, Z. & Shen, Z.-X. Angle-resolved photoemission studies of the cuprate superconductors. *Rev. Mod. Phys.* **75**, 473-541 (2003).

14. Norman, M. R. & Pepin, C. The electron nature of high temperature cuprate superconductors. *Rep. Prog. Phys.* **66**, 1547-1610 (2003).

15. Basov, D. N. & Timusk, T. Electrodynamics of high-$T_c$ superconductors, *Rev. Mod. Phys.* **77**, 721-779 (2005).

16. Zasadzinski, J. F. *et al.* Correlation of tunneling spectra in $Bi_2Sr_2CaCu_2O_{8+\delta}$ with the resonance spin excitation. *Phys. Rev. Lett.* **87**, 067005 (2001).

17. Hwang, J, Timusk, T. & Gu, G. D. High-transition-temperature superconductivity in the absence of the mangetic-resonance mode. *Nature* **427**, 714-717 (2004).

18. Arai, M. *et al.* Incommensurate spin dynamics of underdoped superconductor $YBa_2Cu_3O_{6.7}$. *Phys. Rev. Lett.* **83**, 608-611 (1999).

19. Bourges, P. *et al.* The spin excitation spectrum in superconducting $YBa_2Cu_3O_{6.85}$. *Science* **288**, 1234-1237 (2000).

20. Tranquada, J. M. *et al.* Quantum magnetic excitations from stripes in copper oxide superconductors. *Nature* **429**, 534-538 (2004).

21. Christensen, N. B. *et al.* Dispersive Excitations in the High-Temperature Superconductor $La_{2-x}Sr_xCuO_4$. *Phys. Rev. Lett.* **93**, 147002 (2004).

22. Tranquada, J. M. Neutron scattering studies of antiferromagnetic correlations in cuprates, preprint cond-mat/0512115 at http://xxx.lanl.gov/ (cited 6 Dec. 2005).

23. Dai, P. *et al.* Electronic inhomogeneity and competing phases in electron-doped superconducting $Pr_{0.88}LaCe_{0.12}CuO_{4-\delta}$. *Phys. Rev. B* **71**, 100502(R) (2005).

24. Kang, H. J. *et al.* Electronically competing phases and their magnetic field dependence in electron-doped nonsuperconducting and superconducting $Pr_{0.88}LaCe_{0.12}CuO_{4-\delta}$. *Phys. Rev. B* **71**, 214512 (2005).

25. Yamada, K. *et al.* Commensurate spin dynamics in the superconducting state of an electron-doped cuprate superconductor. *Phys. Rev. Lett.* **90**, 137004 (2003).

26. Fujita, M. *et al.* Magnetic and superconducting phase diagram of electron-doped $Pr_{1-x}LaCe_xCuO_4$. *Phys. Rev. B* **67**, 014514 (2003).

27. Boothroyd, A. T., Doyle, S. M., Paul, McK. D. & Osborn, R. Crystal-field excitations in $Nd_2CuO_4$, $Pr_2CuO_4$, and related *n*-type superconductors. *Phys. Rev. B* **45**, 10075-10086 (1992).

28. Kivelson, S. A. *et al.* How to detect fluctuating stripes in the high-temperature superconductors. *Rev. Mod. Phys.* **75**, 1201-1241 (2003).

29. Aronson, M. C. *et al.* Non-Fermi-Liquid scaling of the magnetic response in $UCu_{5-x}Pd_x$ (*x*=1, 1.5). *Phys. Rev. Lett.* **75**, 725-728 (1995).

30. Matsui, H. *et al.* Direct observation of a nonmonotonic $d_{x^2-y^2}$-wave superconducting gap in the electron-doped high-$T_c$ superconductor $Pr_{0.89}LaCe_{0.11}CuO_4$. *Phys. Rev. Lett.* **95**, 017003 (2005).

**Supplementary Information** is linked to the online version of the paper at www.nature.com/nature.

**Acknowledgements** We thank Elbio Dagotto, Hong Ding, and Shoucheng Zhang for helpful discussions. We also thank Yoichi Ando's group for teaching us how to grow high-quality single crystals of PLCCO. S.D.W. and S. L are supported by the US National Science Foundation. S. C. is supported by the US DOE Division of Materials Science, Basic Energy Sciences. This work is also supported by the US DOE through UT/Battelle LLC.

**Author Information** Reprints and permissions information are available at npg.nature.com/reprintsandpermissions. The authors declare no competing financial interests. Correspondence and requests for materials should be addressed to P.D. (daip@ornl.gov).

**Figure 1 Magnetic susceptibility and a summary of neutron scattering results. a,** Schematic diagrams of real and reciprocal space of the CuO$_2$ with the dashed box showing the first Brillouin zone and the arrow indicating the **Q**-scan direction. Temperature dependence of the magnetic susceptibility for the three crystals (mosaicity < 1°) investigated. Our neutron scattering experiments were performed using pyrolytic graphite (PG) as monochromators and analyzers on the HB-1 triple-axis spectrometer at the High-Flux Isotope Reactor, Oak Ridge National Laboratory, and on the SPINS and BT-9 triple-axis spectrometers at the NIST Center for Neutron Research. We label the momentum transfer **Q** = ($q_x$, $q_y$, $q_z$) as (H, K, L) = ($q_x a/2\pi$, $q_y a/2\pi$, $q_z c/2\pi$) in the reciprocal lattice units (rlu) appropriate for the tetragonal unit cell of PLCCO (a = b = 3.98, and c = 12.27 Å). The three single crystals with a total mass of ~9 grams are co-aligned to within 1° in the [H, H, L] or [H, K, 0] zone. **b,** Elastic scattering along the [0.5, 0.5, L] direction through the (0.5,0.5,1) AF Bragg position at 600 mK and 10 K (ref. 23). The temperature difference spectrum in **c** shows no static AF order at 600 mK. **d,** The full-width at half maximum (FWHM) of the magnetic response with the instrumental resolution marked as horizontal bars. The dashed lines are guides to the eye. **e,** Summary of the resonance energy as a function of $T_c$ for hole-doped YBCO (refs. 3-8), Bi(2212) (refs. 9,10), Tl$_2$Ba$_2$CuO$_{6+\delta}$ (ref. 11), and electron-doped PLCCO. The dashed line is the best fit with $E_r$ = 5.8 $k_B T_c$.

**Figure 2 The wavevector, energy, and temperature dependence of the magnetic scattering around Q = (1/2, 1/2, 0) for 0.5 ≤ $\hbar\omega$ ≤ 4.5 meV.** The

experiments were performed on SPINS with a fixed neutron final energy $E_f$ = 3.7 meV and a cold Be filter before the analyzer. **a-c**, $Q$-scans along the $[H, H, 0]$ direction for $\hbar\omega$ = 0.5, 1.5, and 3.5 meV at $T$ = 2, 30, and 55 K. Center brackets are instrumental resolutions. For $\hbar\omega$ = 0.5 meV, Gaussian fits on linear backgrounds give an amplitude $A$ = 31.3 ± 4.2 counts/10 mins, width $W$ = 0.011 ± 0.0016 rlu, background $Bckg$ = 19.2 ± 1.2 counts/10 mins at 2 K; $A$ = 38.3 ± 4.9 counts/10 mins, $W$ = 0.0104 ± 0.0014 rlu, $Bckg$ = 19.9 ± 1.2 counts/10 mins at 30K. For $\hbar\omega$ = 3.5 meV, $A$ = 19.9 ± 5.2 counts/10 mins, $W$ = 0.098 ± 0.0027 rlu, $Bckg$ = 16.1 ± 1.3 counts/10 mins at 2 K; $A$ = 16.6 ± 4.9 counts/10 mins, $W$ = 0.0094 ± 0.0029 rlu, $Bckg$ = 16.9 ± 1.2 counts/10 mins at 30K. At $\hbar\omega$ = 0.5 meV, the integrated intensities (defined as the sum of raw scanned intensities above the linear fitted background) are 96 ± 15 counts/10 mins at 2 K and 105 ± 14 counts/10 mins at 30 K. At $\hbar\omega$ = 3.5 meV, they are 61 ± 12 counts/10 mins at 2 K and 61 ± 12 counts/10 mins at 30 K. **d,** Temperature dependent scattering at **Q** = (1/2, 1/2, 0) and (0.56, 0.56, 0) for $\hbar\omega$ = 1.5 meV. The background is independent of temperature below ~ 50 K and the magnetic signal becomes much weaker at 80 K. **e,** Constant-**Q** scans at the ridge of magnetic scattering at 2 K, 30 K, and 55 K, compared with the temperature-independent background scattering for 2 ≤ $T$ ≤ 30 K.

**Figure 3 The wavevector and energy dependence of the scattering around Q = (1/2, 1/2, 0) below and above $T_c$.** The experiments were carried out on HB-1 with $E_f$ = 13.5 meV and a PG filter before the analyzer. **a-c,** Q-scans along the [H, H, 0] direction for $\hbar\omega$ = 3.5, 8, and 10 meV at T = 2, 30, and 80 K. The gradual rise in the background scattering with increasing energy (and decreasing scattering angle) is due to the tail of the strong $Pr^{3+}$ crystalline electric field (CEF) excitation at 18 meV and the low-angle background scattering[27]. For $\hbar\omega$ = 10 meV, Gaussian fits on sloped linear backgrounds give amplitudes A = 122 ± 4 counts/10 mins, width W = 0.023 ± 0.001 rlu, background Bckg = 218 ± 9 counts/10 mins at 2 K; A = 84 ± 9 counts/10 mins, W = 0.024 ± 0.004 rlu, Bckg = 212 ± 7 counts/10 mins at 30 K; and A = 87 ± 9 counts/10 mins, W = 0.023 ± 0.004 rlu, Bckg = 200 ± 7 counts/10 mins at 80 K. **d,** Energy scans at **Q** = (1/2, 1/2, 0) where the magnetic scattering peaks, and background scattering at (0.587, 0.587, 0). The maximum energy transfer of 13 meV at **Q** = (1/2, 1/2, 0) is limited by kinematic constraints. The weak dispersion-less peak at 6 meV arises from a previously unknown $Pr^{3+}$ CEF level. Its intensity is temperature independent between 2 K and 30 K within the statistics of our measurements and thus can be regarded as background scattering. **e,** Temperature difference spectrum between 2 K and 30 K suggests a resonance-like enhancement at ~11 meV. See supplementary information for more details on the temperature dependence of the CEF levels.

**Figure 4 The wavevector, energy, and temperature dependence of the scattering around Q = (1/2, 1/2, 0).** Data in **a-c** are collected using BT-9 with $E_f$ = 28 meV and a PG filter before the analyzer. This geometry allows the kinematic constraints to be satisfied at (1/2, 1/2, 0) for $\hbar\omega \leq 16$ meV. **a,** The energy scans along the ridge of magnetic scattering at (1/2, 1/2, 0) were counted for ~2 hours/point to obtain the statistics shown. The background scattering at (0.6, 0.6, 0) was counted ~30 minutes/point and showed no observable difference between 2 K and 30 K. **b,** Wavevector scans along the [H, H, 0] direction around (1/2, 1/2, 0) at $\hbar\omega$ = 10 and 15 meV. The kinematic constraints allow only half of the Q-scan at $\hbar\omega$ = 15 meV. **c,** Temperature difference (2 K - 30 K) spectrum at **Q** = (1/2, 1/2, 0) shows the resolution-limited resonance at $\hbar\omega$ =11 meV. The energy resolution of the spectrometer is ~3.7 meV in FWHM at $\hbar\omega$ = 10 meV. **d,** Black squares show temperature dependence of the neutron intensity (~1 hour/point) at (1/2, 1/2, 0) and 10 meV obtained on HB-1 (Fig. 3). Green diamonds are integrated intensity of the localized signal centered around **Q** = (1/2, 1/2, 0) above backgrounds in Fig. 3c. The line is a guide to the eye. **e,** Q-scans at $\hbar\omega$ = 10 meV obtained on BT-9 with $E_f$=14.7 meV and collimations 40´-40´-40´-80´. The Gaussian fits have A = 25.4 ± 3.4 counts/10 mins, W=0.022 ± 0.004 rlu, *Bckg* = 80 ± 3 counts/10 mins at 2 K and A = 15.2 ± 2.7 counts/10 mins, W = 0.023 ± 0.005 rlu, *Bckg* = 78 ± 2 counts/10 mins at 30 K.

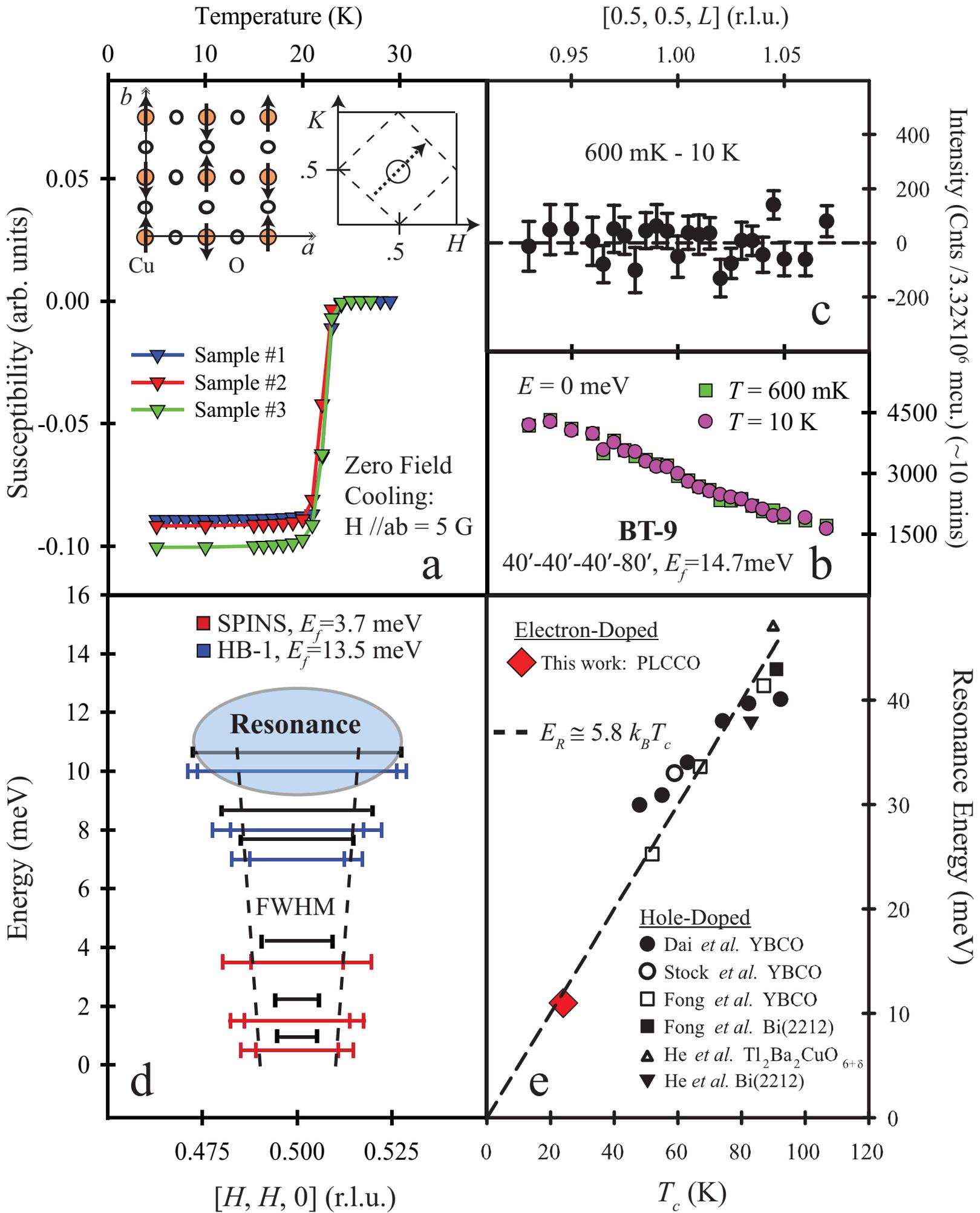

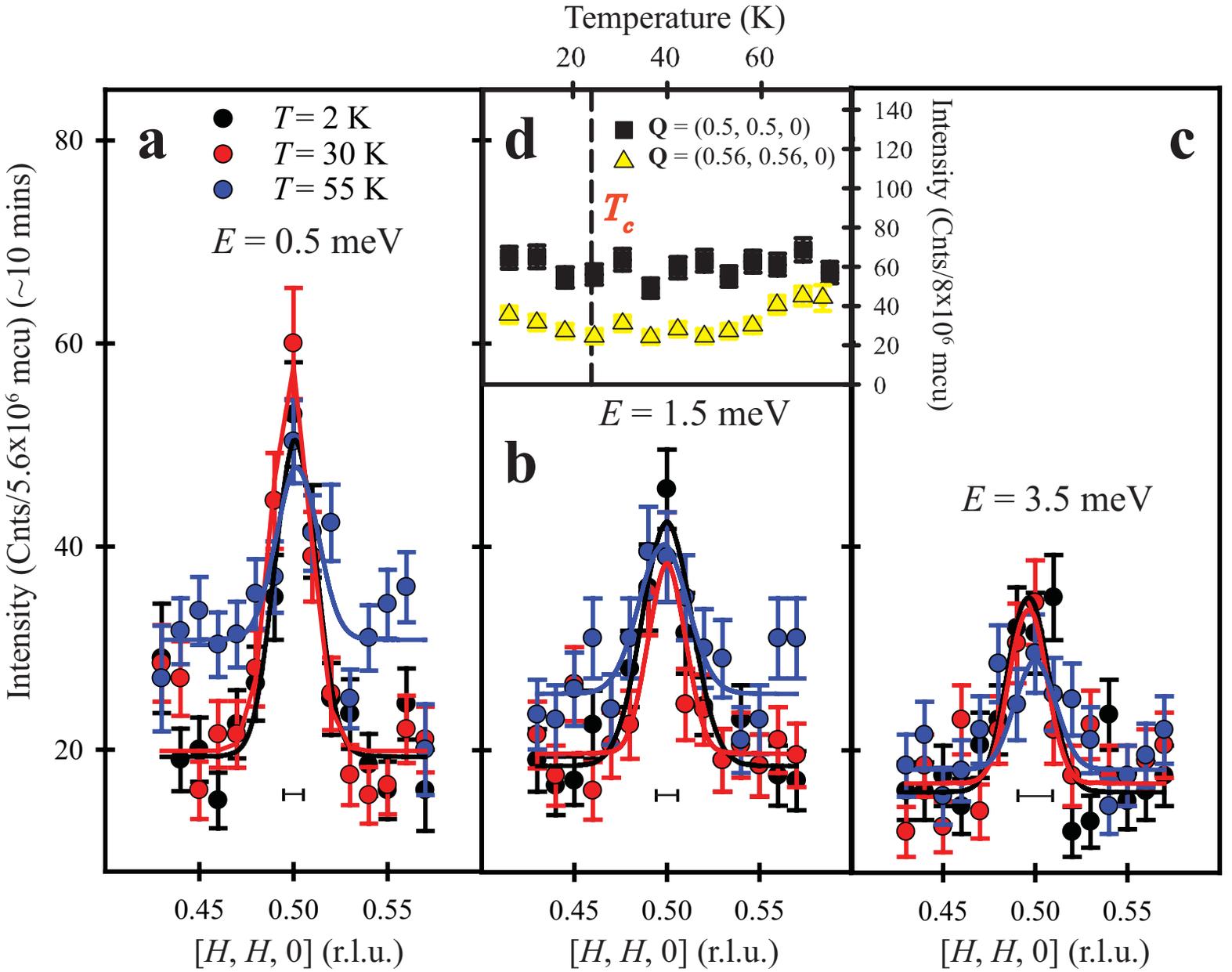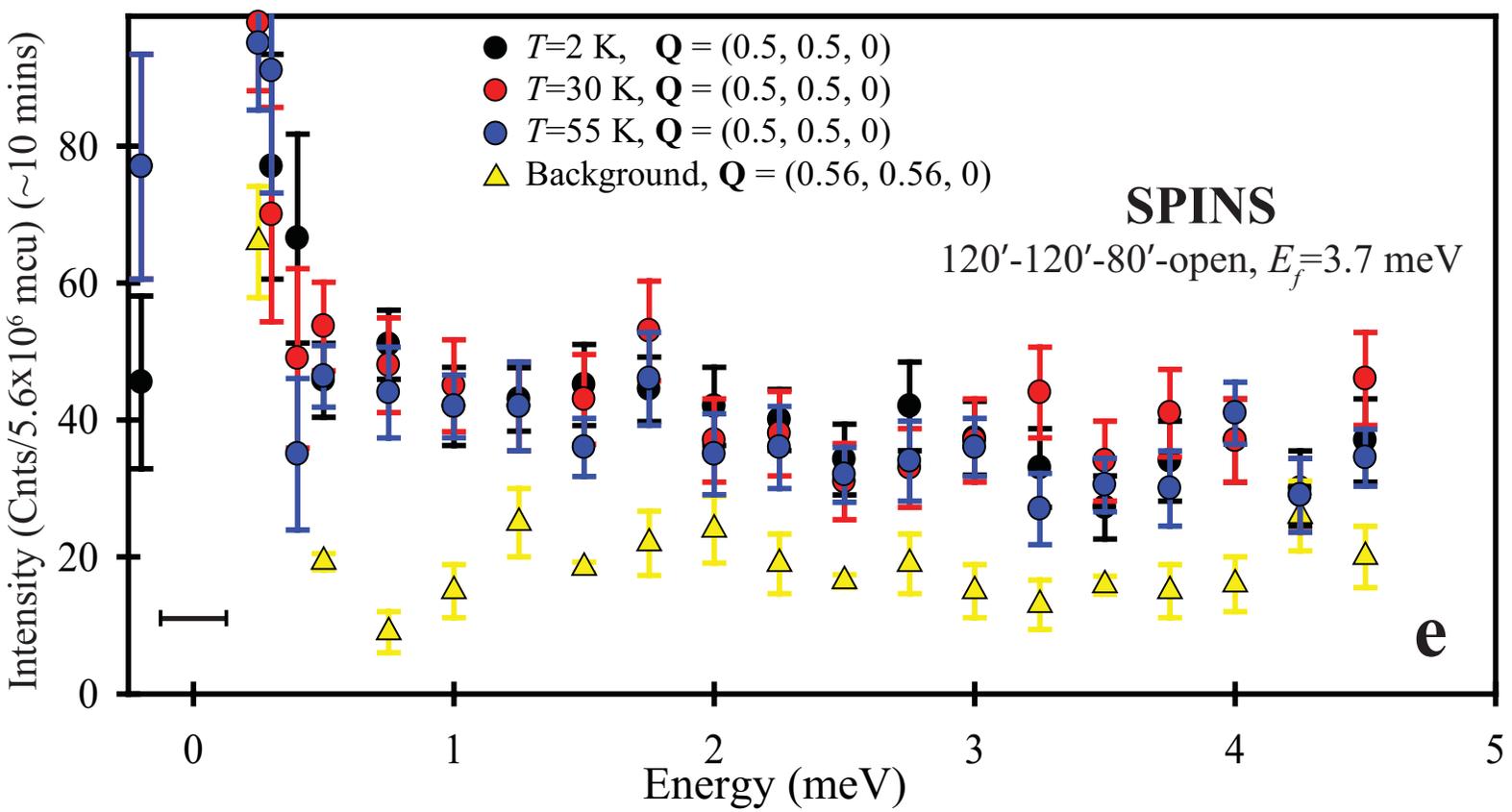

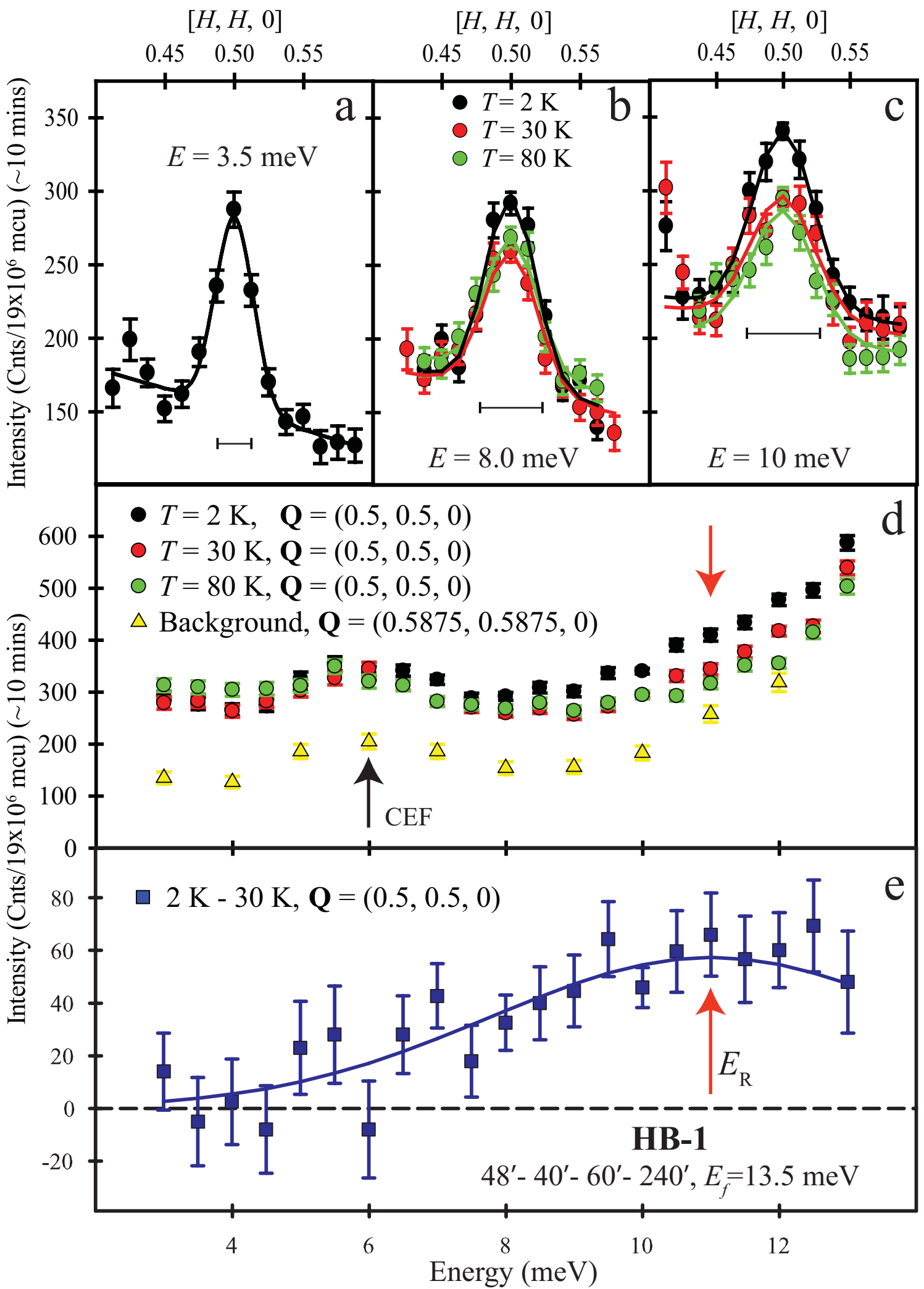

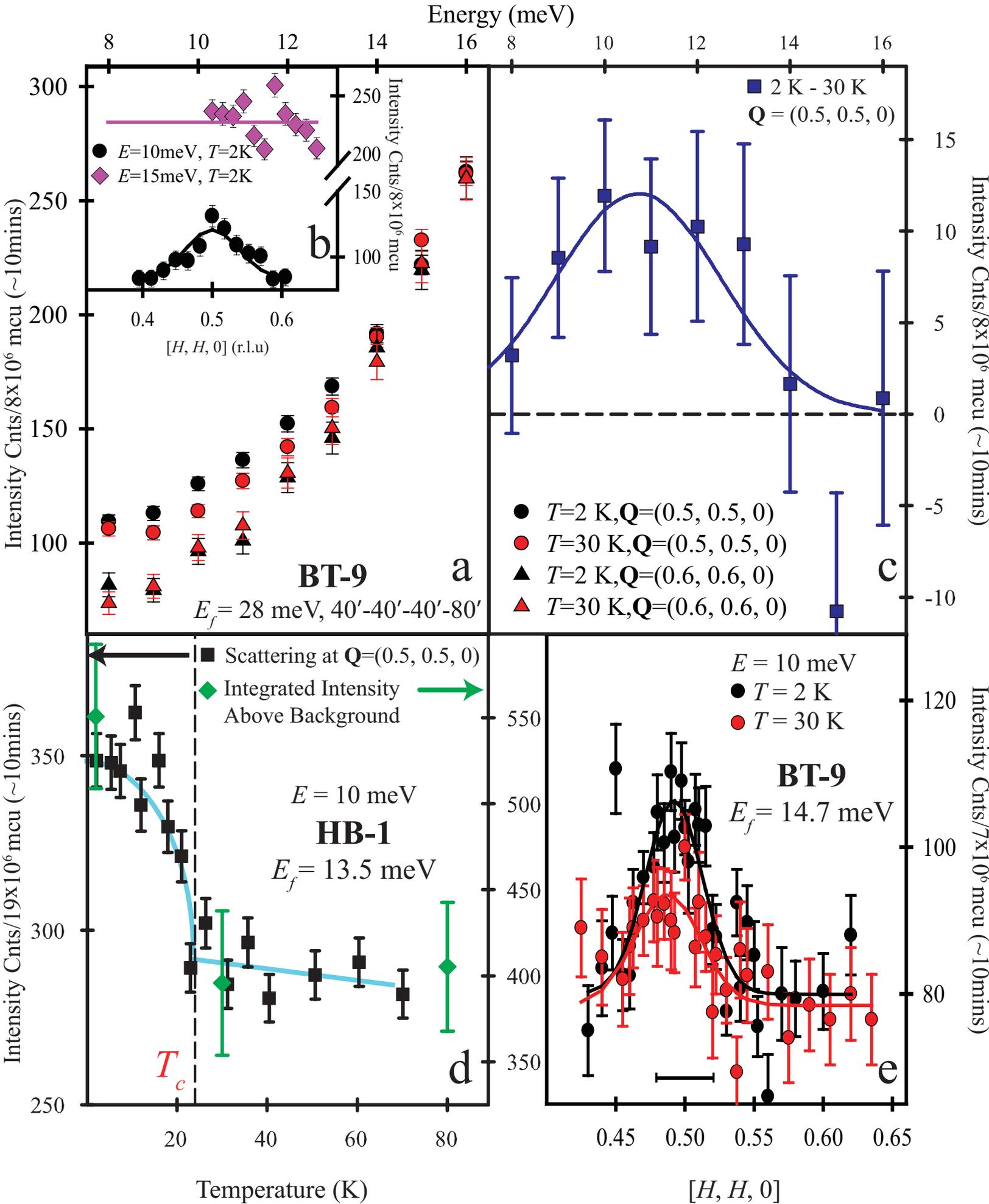